\documentclass[fleqn,usenatbib]{mnras}

\usepackage{newtxtext,newtxmath}

\usepackage{soul}

\usepackage[T1]{fontenc}

\DeclareRobustCommand{\VAN}[3]{#2}
\let\VANthebibliography\thebibliography
\def\thebibliography{\DeclareRobustCommand{\VAN}[3]{##3}\VANthebibliography}

\usepackage{graphicx}	
\usepackage{amsmath}	

\title[A millimetre-wave hyper-spectral device]{A millimetre-wave superconducting hyper-spectral device}

\author[U. Chowdhury et al.]{
U.~Chowdhury,$^{1,2}$
M.~Calvo,$^{1,2}$
J.~Goupy,$^{1,2}$
F.~Levy-Bertrand,$^{1,2}$
and A.~Monfardini$^{1,2}$\thanks{E-mail: monfardini@neel.cnrs.fr}
\\
$^{1}$Univ. Grenoble Alpes, CNRS, Grenoble INP, Institut N\'eel, 38000 Grenoble, France\\
$^{2}$	Groupement d'Int\'er\^et Scientifique KID, 38000 Grenoble and 38400 Saint Martin d'H\`eres, France\\
}

\date{Accepted 03/08/2023. Received 04/07/2023; in original form 01/01/2023}

\pubyear{2015}

\begin{document}
\label{firstpage}
\pagerange{\pageref{firstpage}--\pageref{lastpage}}
\maketitle

\begin{abstract}
Millimetre-wave observations represent an important tool for Cosmology studies. The Line Intensity Mapping (LIM) technique has been proposed to map in three dimensions the specific intensity due to line (e.g. [CII], CO) emission, for example from the primordial galaxies, as a function of redshift. Hyper-spectral integrated devices have the potential to replace the current Fourier transform, or the planned Fabry-Perot-based instruments operating at millimetre and sub-millimetre wavelengths. The aim is to perform hyper-spectral mapping, with a spectral resolution $R = \lambda / \Delta\lambda = 100 - 1000$, over large, i.e. thousands of beams, instantaneous patches of the Sky. The innovative integrated device that we have developed allows avoiding moving parts, complicated and/or dispersive optics or tunable filters to be operated at cryogenic temperatures. The prototype hyper-spectral focal plane is sensitive in the 75-90~GHz range and contains nineteen horns for sixteen spectral-imaging channels, each selecting a frequency band of about 0.1~GHz. For each channel a conical horn antenna, coupled to a planar superconducting resonant absorber made of thin aluminium, collects the radiation. A capacitively coupled titanium-aluminium bilayer Lumped Element Kinetic Inductance Detector (LEKID) is then in charge of dissipating and sensing the super-current established in the resonant absorber. The prototype is fabricated with only two photo-lithography steps over a commercial mono-crystalline sapphire substrate. It exhibits a spectral resolution $R = \lambda / \Delta\lambda \approx 800$. The optical noise equivalent power of the best channels is in the observational relevant $4\cdot10^{-17}~W/\sqrt{Hz}$ range. The average sensitivity of all the channels is around $1\cdot10^{-16}~W/\sqrt{Hz}$. The device, as expected from 3-D simulations, is polarisation-sensitive, paving the way to spectro-polarimetry measurements over very large instantaneous field-of-views.
\end{abstract}

\begin{keywords}
Instrumentation -- Detectors -- Line Intensity Mapping
\end{keywords}



\section{Introduction}

The Line Intensity Mapping (LIM) technique (\cite{LIM2017}, \cite{LIM2019}) aims to map the specific intensity due to line (e.g. [CII], CO) emission. Such a mapping, if achieved with spectrally-resolving instruments, allows to disentangle the line emission as a function of the redshift. The end result is thus a three-dimensional view of the observed region. LIM observations are typically planned on single dish telescopes, since the goal is not to resolve the individual objects but on the opposite to maximize the mapping speed, i.e. enhancing the sensitivity and the field-of-view (FoV). 

One possible technique is to use low-temperature on-chip spectrometers (\cite{microspec}, \cite{SPT_SLIM}, \cite{WSPEC},  \cite{SuperSpec}, \cite{DESHIMA}, \cite{OMKID}), aiming to achieve intermediate resolution, i.e. $R = \lambda / \Delta\lambda = 100 - 1000$. The main limitation is, despite the recent dramatic improvements, the finite number of available readout electronics channels. In the near future, it will not be possible to cover, with this technique, the large instantaneous FoV that are required for the planned LIM experiments. On top of that, for technological reasons the existing devices exhibit relatively low overall quantum efficiency. Similar spectral resolutions, over large FoV, are currently achieved using FTS (Fourier transform spectrometers) \citep{concerto2020, concerto2021}. The drawback is in this case represented by the higher optical background per pixel, translating into lower per-pixel sensitivity. Furthermore, the optical design is complicated, requiring for example large moving mirrors at room temperature. 

An efficient alternative that has been proposed is the use of Fabry-Perot elements operating at low temperatures. This is for example planned for the EoR-Spec project \citep{fabry-perot}. In this case the observing frequency is swept by the settable narrow-band Fabry-Perot filter to reconstruct the spectrum. This approach is compatible with large FoV, i.e. thousands spatial pixels simultaneously on the Sky, and does not suffer from the high per-pixel load limitation. The drawbacks lies in the narrowness of the instantaneous bandwidth and the stringent mechanical requirements to scan and keep the mirrors parallel.

In this work we present the design, fabrication and optical characterisation of an innovative hyper-spectral device based on Lumped Element Kinetic Inductance Detectors (LEKID). Our HYPKID prototype focal plane targets the so-called "3-mm atmospheric window". The extension to higher frequencies, e.g. 2-mm and 1-mm windows, is feasible.

The HYPKID fabrication process is made by only two photo-lithographic steps and does not require the deposition of dielectrics. The incoming signal, at a frequency between 75~GHz and 90~GHz, is directly absorbed by superconducting c-shaped resonators. This direct absorption is a major difference of concept from the previously proposed on-chip spectrometers in which photons are guided from an antenna through a planar transmission line up to superconducting resonant filters. The removal of the transmission line allows a gain in the overall quantum efficiency. 

In terms of focal plane efficiency, compatibility with large FoV, photon noise and mapping speed, HYPKID is equivalent to a Fabry-Perot instrument, with similar drawback of narrow instantaneous bandwidth.The main advantages of the HYPKID when compared to existing technologies are: a) the easier detectors fabrication, comparable to the well-proven LEKID imaging arrays \citep{nika2}; b) the higher quantum efficiency with respect to the existing on-chip spectrometers; c) the easy optical coupling, in particular in comparison to the MpI instruments.

\section{Design and fabrication}

\begin{figure*}[ht]
\begin{center}
\resizebox{\linewidth}{!}{\includegraphics{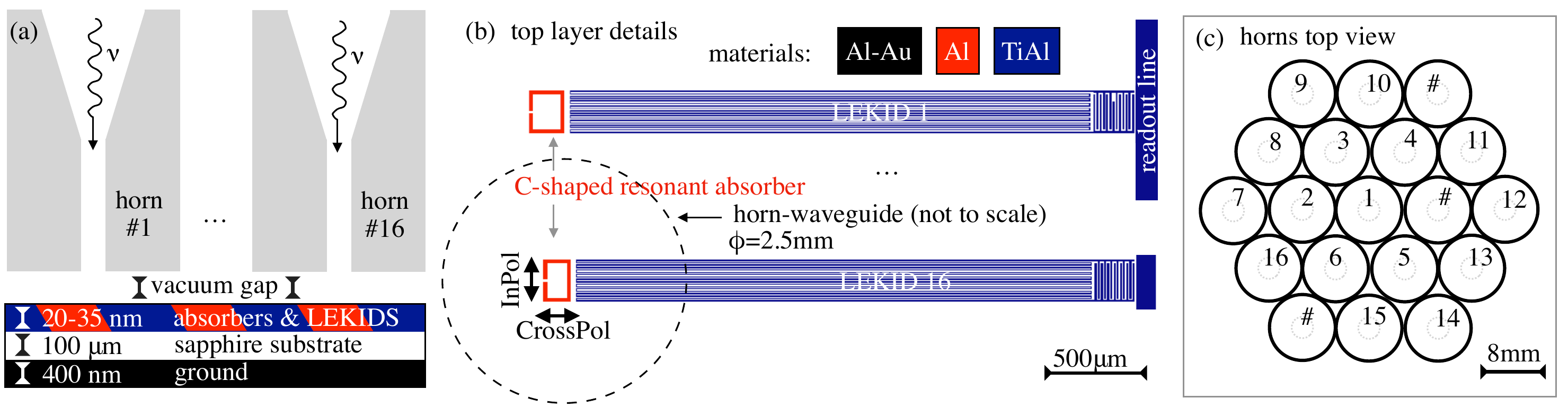}}
\caption{\textbf{HYPKID device schematic.} The device targets the 75-90~GHz range with 16 imaging/spectral channels (out of 19 horns). (a) Side view. The radiation is collected through a horn-waveguide radiating at the end toward the patterned superconducting layers level, realised on top of the sapphire substrate. The ground plane is on the other side of the substrate. (b) Top view. The polarised incoming radiation (InPol or CrossPol) excites a resonance in the superconducting c-shape spectral-selective absorbers. The filtered signal is dissipated by the LEKID made of TiAl (spectroscopic gap around 70~GHz). The resonance frequencies of the LEKID are monitored with a GHz readout line. (c) The hexagonal-packaged 19-horns device. Numbers indicate the position of the sixteen channels. Each channel detects a different wavelength. The three \# symbols indicate the position of LEKIDs without c-shape resonant absorber. These LEKIDs aim to measure the continuum background.}
\label{fig_design}
\end{center}
\end{figure*}

Our prototype HYPKID device has been designed for and fabricated on a two inches 110~$\mu m$-thick monocrystalline C-plane sapphire substrate. It contains nineteen horns; sixteen of them with spectral capabilities which will refered as \textit{spectral channels} afterwards and three, LEKIDs without any absorber for continuum measurement. A schematic view is presented in figure~\ref{fig_design}.

The incoming signal is collected by an array of nineteen smooth conical horn antennas disposed over an hexagonal pattern. The 12~mm long horn ends up in a portion of cylindrical waveguide with a diameter of 2.5~mm. The cutoff frequency is around 70~GHz, and the antenna is expected to be single-mode ($TE_{11}$) until the onset of the $TM_{01}$ mode above 91~GHz. For sixteen horns, the waveguide itself radiates toward the resonant absorbers that provide the spectral capabilities. The sixteen absorbers are c-shaped half wave planar resonators whose length is varied to cover the range of 75-90~GHz. The spectral resolution of each filter is set by its coupling quality factor which is adjusted via the distance to the LEKID to target a bandwidth of 0.1-0.2~GHz. The filters shape aims at maximizing the current next to the LEKID, and optimising the dissipation efficiency in the detector. Thus, the mm wave signal is absorbed in the LEKIDs where it generates quasiparticles. The vacuum gap between the horns and the on chip device is of approximately 700~$\mu$m, to be further optimised for the next generation of devices.

The resonance frequencies of the LEKID are affected by the change of their kinetic inductance. A single GHz (microstrip) readout line is frequency-multiplexing the sixteen channels on a single electronics IN/OUT pair of coaxial cables. The readout rate for the HYPKID is fixed at 46~Hz.

\subsection{Design}
\label{sec:design}

\begin{figure}
\begin{center}
\resizebox{\linewidth}{!}{\includegraphics{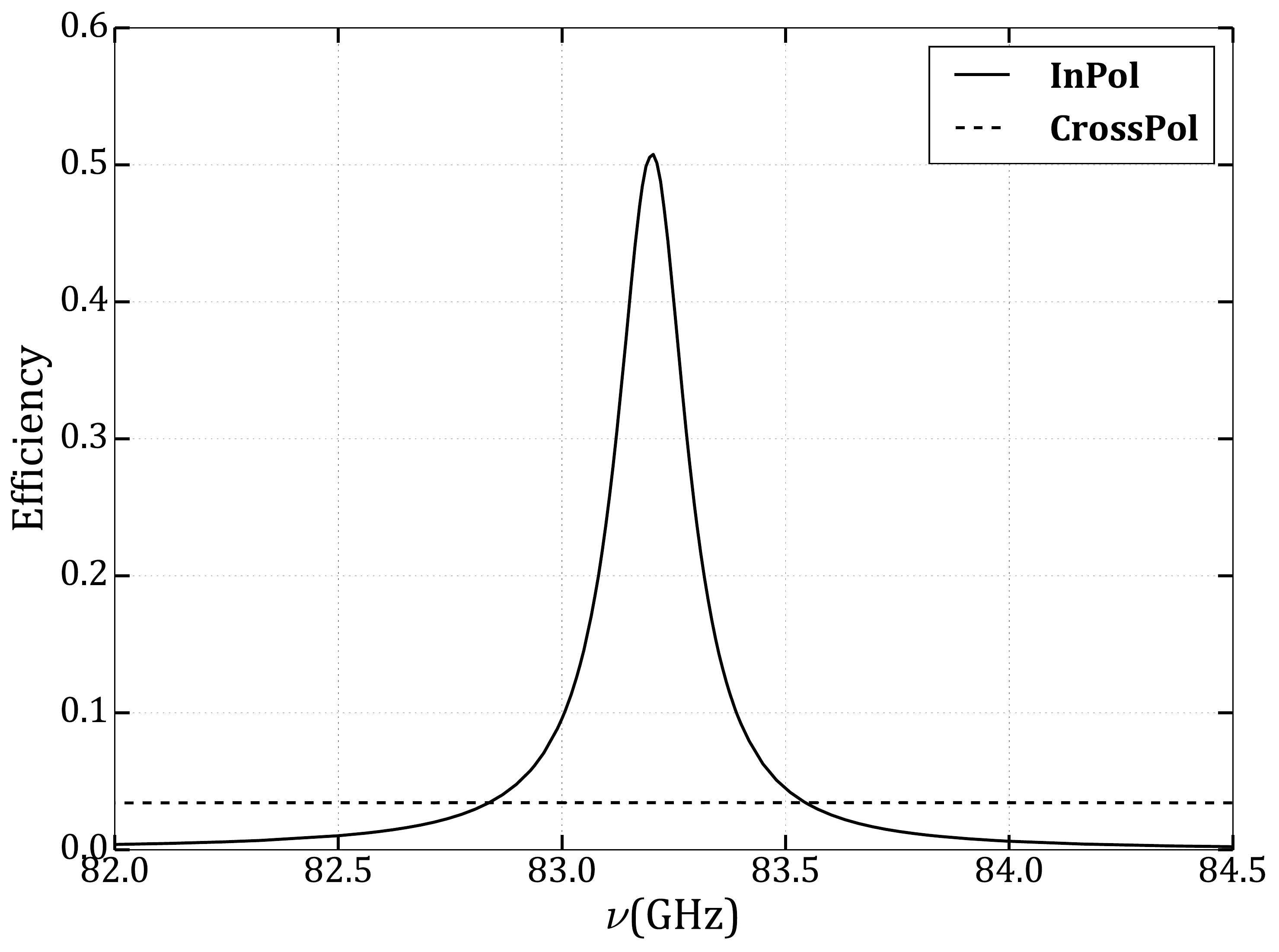}}
\caption{\textbf{Simulated response of a channel.} In the simulation we include the cylindrical waveguide terminating the horn, the vacuum gap, the resonant absorber, the LEKID, the dielectric substrate and the ground plane. At the peak of the absorber resonance the fraction of incoming power that is dissipated and contributes to the signal in the LEKID} is of around 50\%. The FWHM of the resonance is compatible with a spectral resolution R = 450. The predicted cross-polarisation signal is around 5\%.
\label{fig_1D_CST}
\end{center}
\end{figure}

To establish the possible use of the c-shaped millimeter-wave planar absorber as efficient termination for the horn-waveguide we have simulated the structure using CST Studio Suite\textregistered. The Titanium-Aluminium LEKID is defined in the simulation as an Ohmic sheet with a resistance of 2~$\Omega$ per square. The aluminium absorber on the other hand has a purely imaginary impedance of 1~pH per square to reproduce the kinetic inductance. In a first simulation (figure \ref{fig_1D_CST}) including the cylindrical waveguide, the vacuum gap, the resonant absorber, the LEKID, the dielectric substrate and the ground plane, we have calculated the fraction of the accepted power through the waveguide dissipated into the lossy LEKID.  At the absorber resonance peak and for the current prototype device, we expect an efficiency of around 50\% for the InPol polarization. This confirms that the device that we propose is an efficient option, especially compared to the existing on-chip spectrometers that exhibit lower quantum efficiencies. The cross-polarised signal is predicted at around 5\%. Our simulations suggest that the efficiency can be increased up to 80\% (for the InPol polarization). To achieve this efficiency the key parameter to adjust is the vacuum gap distance between the end of the waveguide and the superconducting resonant absorber. This distance, in our prototype, was fixed due to unavoidable mechanical constrains in the sample holder. 

\begin{figure}
\begin{center}
\resizebox{\linewidth}{!}{\includegraphics{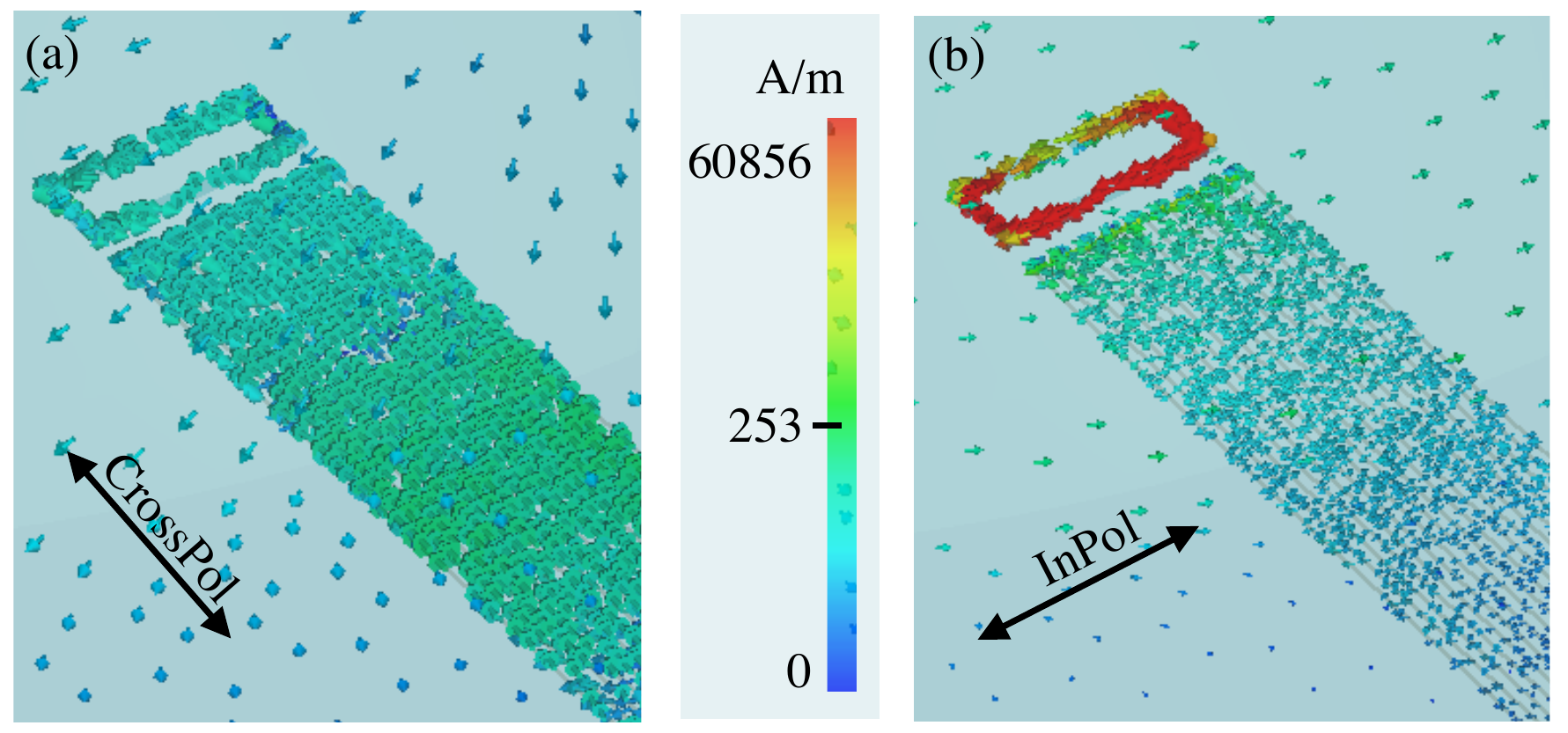}}
\caption{\textbf{3D electromagnetic simulation.} Distribution of the surface current for CrossPol (left panel) and InPol (right panel) illumination through the waveguide. The current is calculated at the superconducting absorber resonance frequency, i.e. 88~GHz. It is evident that the radiation collection is achieved in the absorber for the InPol case, while it is mostly happening directly in the LEKID inductive meander for the CrossPol situation.}
\label{fig_CST}
\end{center}
\end{figure}

The figure \ref{fig_CST} shows the simulated surface current in the metal layers, i.e. LEKID and c-shaped absorber. The plots evidence the polarisation selectivity of the device. The c-shape planar structure absorbs mostly the InPol direction of the electric field. The cross-polarised signal is mainly due to the direct absorption into the LEKID inductive meander that is exposed to the waveguide output. 

The resonant frequency of each LEKID is tuned by adjusting the length of the meander (L) and one of the fingers of the capacitor (C). The resonance frequencies range between 1.5~GHz and 1.8~GHz. The internal quality factor $Q_i$ (material) and the coupling quality factor $Q_c$ (design) are both in the $10^5$ range. The LEKID, as in our standard imaging devices \citep{nika2}, are coupled to a 50~$\Omega$ micro-strip readout-line.

\subsection{Fabrication}
\label{sec:fabrication}

Three types of superconducting materials have been used to fabricate the HYPKID: pure aluminum (20~nm), titanium-aluminum bi-layer (10~nm~/~25~nm) and aluminum cover with gold (200~nm~/~200~nm). The pure aluminum superconductor (Al) is a lossless conductor for frequencies smaller than its spectroscopy gap $2\Delta_{Al}/h\sim$110~GHz. It is used for the lossless c-shaped superconducting absorbers and patterned via standard UV lithography and lift-off. For the LEKIDs, we use the titanium-aluminum bi-layer (TiAl) as this layer dissipates the radiation for frequencies higher than its spectroscopic gap $2\Delta_{TiAl}/h\sim$70~GHz \citep{catalano2015}. The TiAl bi-layer is also used for the GHz microstrip readout line. This layer is patterned using standard UV lithography, followed by 5:1:1 H$_2$O:NH$_4$OH:H$_2$O$_2$ solution. The aluminum covered with gold (Al-Au) is used for the ground plane on the back-side of the substrate. The gold layer is crucial to ensure a good thermalisation of the device (and thus a fast time constant), whereas the thick aluminum ensures a high-quality (superconducting) electrical ground plane.

\section{Experimental set-up}
The hyper-spectral device is mounted in a custom optical dilution refrigerator (mm-wave camera) with a base temperature of 60~mK. The camera is directly derived from the NIKA (N\'eel IRAM KID Arrays) instrument \citep{monfardini2011}, and also described in \cite{OMKID}.

To characterize the spectral width of the channels, we have used a commercial mm-wave source module coupled to a rectangular emission horn. The mm-wave source is polarised and can be adjusted in the range 75-110~GHz with Hz-like frequency precision. The output power of the source is around 8 mW. It is however not calibrated in intensity. In order to attenuate the signal entering the cryostat, we place five Eccosorb$^{TM}$ sheets in front of the input lens. Each sheet attenuates the signal by a factor around 25. The total attenuation is thus estimated around $10^7$. 

For the estimation of the Noise Equivalent Power (NEP) we use well-calibrated sources: a room-temperature black body (T$_{bb}\sim~$300~K) and a mirror acting as black body with an effective temperature determined by our Sky Simulator \citep{monfardini2011} to be close to "0~K" (T$_{bb}\sim~$0~K).

The signal per channel is defined as the frequency shift of the corresponding LEKID. Each channel corresponds to one pixel pointing on the sky and detects one specific wavelength. In previous publications, we have demonstrated the linear proportionality between the absorbed power and the frequency shift (\cite{swenson2010}, \cite{Calvo2013}). The frequency shift of the LEKID is measured using a dedicated multiplexing electronics \citep{bourrion2013}, synchronized with the mm-source.

\section{Results and discussion}

\begin{figure}
\begin{center}
\resizebox{\linewidth}{!}{\includegraphics{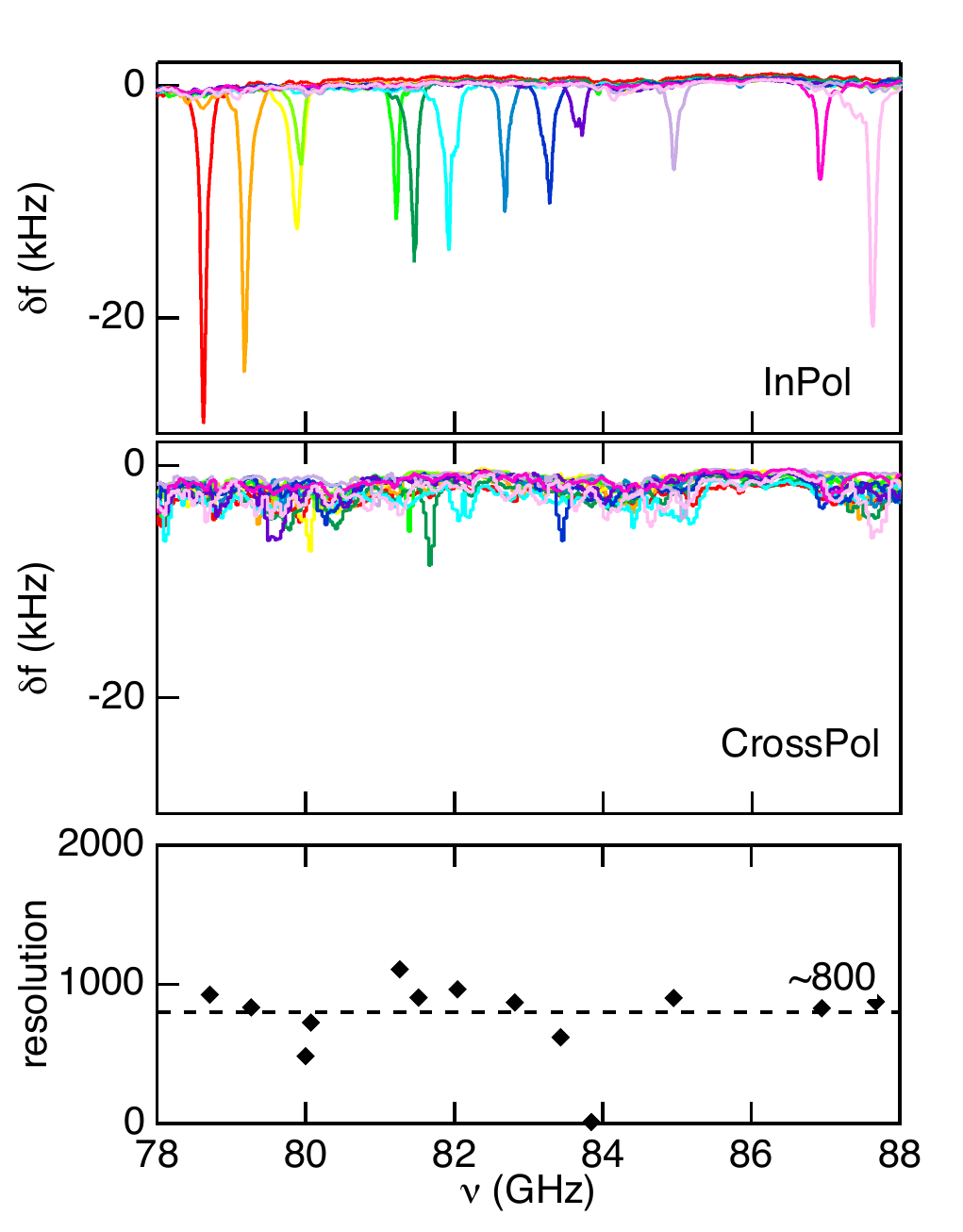}}
\caption{\textbf{HYPKID pixels responses.} Frequency shift of the LEKID as a function of the incoming radiation frequency:  (top) for a polarisation of the mm-source perpendicular to the LEKID inductance lines - InPol; (middle) for the polarisation parallel to the LEKID lines - CrossPol; (bottom) Spectral resolution per channel.}
\label{fig_channels}
\end{center}
\end{figure}

We show in figure~\ref{fig_channels} the HYPKID channels responses and the corresponding spectral resolution measured with the mm-wave source. Thirteen out of the sixteen designed channels are functional, showing a very clear peaked spectral response in the range 78-88~GHz. The origin of the scatter in the response amplitude from channel to channel is not firmly established. Possible origins are power variation of the mm-source throughout the frequency-band, non-uniform illumination of the focal plane and scattering of the coupling among the channels.

As expected from the simulations described above, the device is polarisation selective. The un-filtered background is clearly detected in the CrossPol direction at an amplitude of around 2~kHz, while it is much less pronounced in the InPol direction.
For this polarisation direction, the integrated un-filtered background contributes less to the signal. This is achieved thanks to the orthogonal polarisation absorption of the superconducting absorber and the LEKID.

The measured spectral resolution of our prototype HYPKID, shown on a pixel-per-pixel base in the figure \ref{fig_channels}, is around $R\approx800$. This value is slightly higher than the simulation prediction ($R\approx450$). This might be related to the specific fabrication process used for this first prototype device.

\begin{figure}
\begin{center}
\resizebox{\linewidth}
{!}{\includegraphics{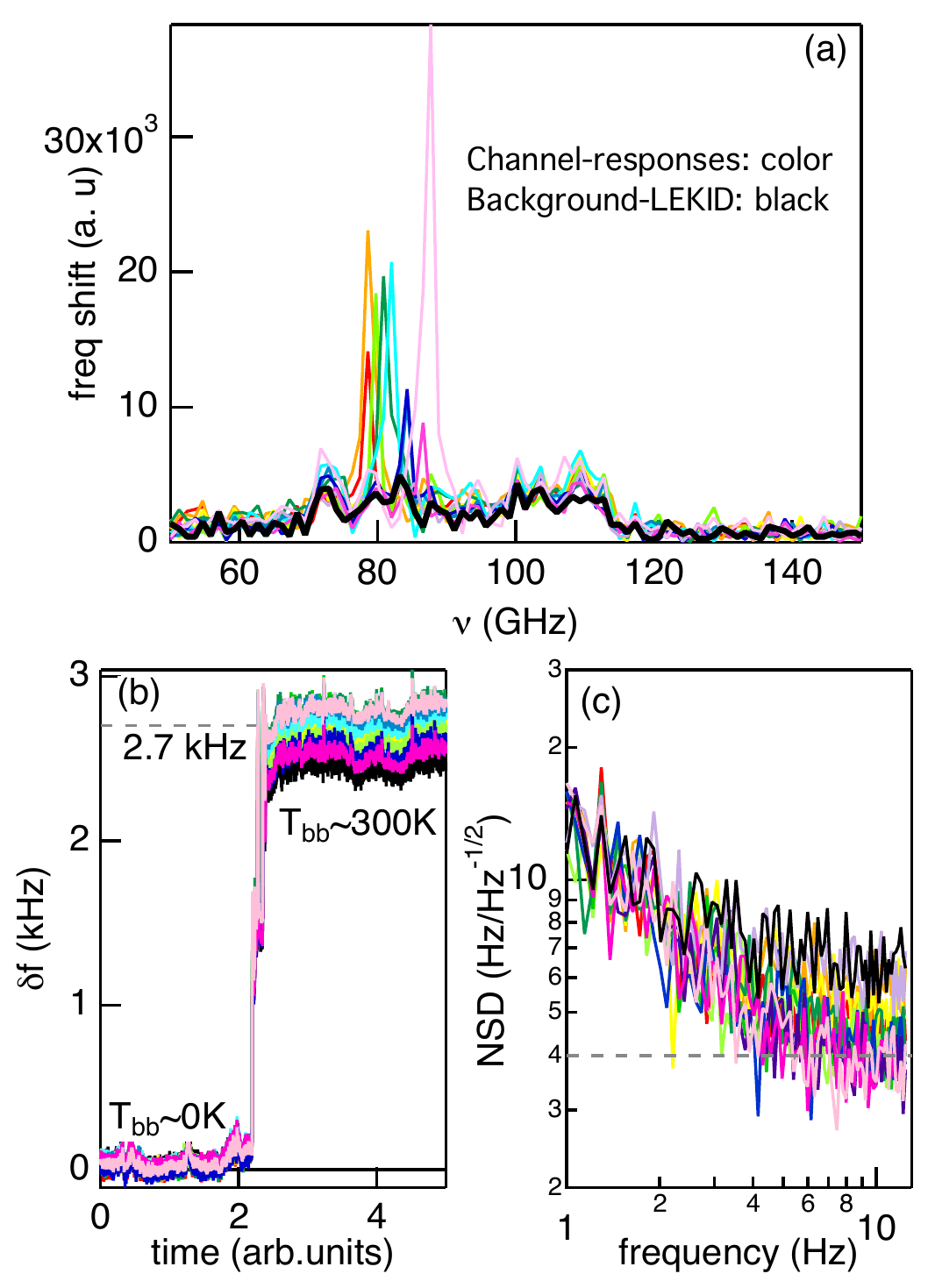}}
\caption{\textbf{Response and noise spectral density of the HYPKID channels.} (a) Martin-Puplett spectral measurement of the functional channels. (b) Frequency shift to a change of optical load: a black body source temperature (T$_{bb}$) from about "0~K" to 300~K. The frequency shift  is of the order of 2.7~kHz. (c) Noise Spectral Density (NSD). At 10~Hz the noise is of the order of 4~$Hz/\sqrt{Hz}$.}
\label{fig_NEP}
\end{center}
\end{figure}

The Noise Equivalent Power (NEP) corresponds to the power producing a signal-to-noise ratio of one in one Hz output bandwidth. It is calculated as: 

\begin{equation}
    \textrm{NEP}=\frac{\Delta W_{opt} \cdot N}{\delta f}
	\label{eq_NEP}
\end{equation}
where $\delta f $ is the frequency shift (Hz) of the LEKID resonance generated by the change of the optical power load $\Delta W_{opt}$ (the power per pixel) and $N$ is the noise spectral density expressed in $Hz/\sqrt{Hz}$. 
Figure~\ref{fig_NEP} shows the measurements required  to evaluate the NEP. 
These measurements were performed with black-body sources and with a polarizer filtering the CrossPol polarisation to reduce further the background contribution. Panel (a) displays the broad band responses of the LEKIDs with a resolution of about 1~GHz obtained with a Martin-Puplett Fourier Transform spectrometer (more details in supplementary figure~3 of reference~\cite{maleeva_circuit_2018}). From the comparison of the background and the channel responses, we estimate that the channel response contributes, in average, to 10\% of the total response. In other words, the standard pixels, i.e. equipped with resonant superconducting absorbers, exhibit on average 10\% more signal with respect to those that only have LEKID (filter-less). Panel (b) shows the frequency shift of the channels for a change of optical load due to a black body source temperature in between "0~K" to 300~K. The frequency shift is of the order of 2.7~kHz. The frequency shift due to the channel response is thus of the order 270~Hz. Panel (c) displays the spectral noise density of the channels at "0~K" optical loading. The noise is the same, within statistical errors, for the case of 300~K. At 10~Hz the noise is of the order of 4~$Hz/\sqrt{Hz}$. The optical power load is evaluated with a three dimensional ray tracing software where the inputs are the spectral luminescence of the black-body source, the optics of the cryostat and the effective collecting surface of the horn. It is assumed that the effective surface area of the horn is 30~mm$^2$ that is 60\% of its physical surface. At 80~GHz, for a 0.1~GHz single polarization band, the optical power load is around 7~fW. The estimated NEP ranges from $4\times10^{-17}W/\sqrt{Hz}$ to $3\times10^{-16}W/\sqrt{Hz}$ , with an average value of $1\times10^{-16}W/\sqrt{Hz}$.    



\section{Conclusions}

We have designed, fabricated and tested a new hyper-spectral imaging device operating in the astronomically relevant millimetre-wavelength band.  The major difference of the concept from the existing on-chip spectrometers is the direct absorption of light in the filterbank, removing elements such as the transmission line that reduce the quantum efficiency. Our device is characterised by a straightforward design and fabrication process, readily compatible with large field-of-view, i.e. thousands beams (pixels), arrays.

The first prototype confirmed the soundness of this innovative device, showing good performance in terms of spectral resolution ($R \approx 800$), polarisation selection and optical sensitivity, i.e. NEP for the best channels in the $4\cdot10^{-17}W/\sqrt{Hz}$ range. The direct comparison of the performance with our on-chip spectrometer \citep{OMKID}, operating in the same band and characterised in the same way, confirms the gain in quantum efficiency as predicted by the electro-magnetic simulations. 
 
The optical sensitivity figures are already interesting for the HYPKID to be considered for the next generation of experiments targeting Line Intensity Mapping observations.

The device being polarisation-sensitive, the hyper-spectral properties can be combined to polarisation measurements. For example, by introducing a polarisation modulator in the beam, i.e. a fast rotating half-wave plate \citep{Pisano2022}. A 45-degrees polariser (splitter) can be inserted to illuminate simultaneously two polarisation-sensitive focal planes in a very compact configuration. 

We plan to build on the promising results presented in this paper and elaborate further, more optimised hyper-spectral devices on different mono-crystalline dielectrics operating up to 300~GHz. We have designed and realised a first 301-horns prototype operating at 150~GHz. Further details will be given in future publications. 

\section*{Acknowledgements}

We acknowledge the specific contribution of the engineer G.~Garde to the device holder and the overall support of the Cryogenics, Electronics and Nanofab groups at Institut N\'eel and LPSC. 
The fabrication of the device described in this paper was conducted at the PTA Grenoble micro-fabrication facility. 
This work has been partially supported by the French National Research Agency through Grant No. ANR-16-CE30-0019 ELODIS2, the LabEx FOCUS through Grant No. ANR-11-LABX-0013, the EU Horizon 2020 research and innovation program under Grant Agreement No. 800923 (SUPERTED) and Grant Agreement No. 78821 (CONCERTO).

\section*{Data Availability}
No astronomical data has been collected for this work. The laboratory data are readily available upon request to the corresponding author.



\bibliographystyle{mnras}
\bibliography{biblio} 








\bsp	
\label{lastpage}
\end{document}